\documentclass[%
 reprint,
superscriptaddress,
 amsmath,amssymb,
 aps,prb,citeautoscript
]{revtex4-1}

\usepackage{graphicx}
\usepackage{dcolumn}
\usepackage{bm}
\usepackage{color}	
\usepackage{hyperref}
\usepackage{tikz}
\usepackage{float}

\newcommand{\minutes}{\,\mathrm{min}}

\newcommand{\degC}{^\circ \mathrm{C}}
\newcommand{\degCmin}{^\circ \mathrm{C/min}}

\newcommand{\s}{\,\mathrm{s}}
\newcommand{\invs}{\,\mathrm{s^{-1}}}

\newcommand{\mm}{\,\mathrm{mm}}

\newcommand{\micron}{\,\mathrm{\mu m}}

\newcommand{\de}{\partial}
\newcommand{\di}{\mathrm{d}}

\newcommand{\gd}{\dot{\gamma}}

\usepackage{tikz}
\newcommand*\circled[1]{\protect\tikz[baseline=(char.base)]{
		\protect\node[shape=circle,draw,inner sep=0.3pt] (char) {#1};}}

\newcommand{\regI}{{\small \emph{i}}}
\newcommand{\regII}{{\small \emph{ii}}}
\newcommand{\regIII}{{\small \emph{iii}}}

\newcommand{\tenpc}{10~\text{\% wt.}}
\newcommand{\fifteenpc}{15~\text{\% wt.}}

\begin{document}

\title{Interplay between wall slip and shear banding in a thixotropic yield stress fluid}

\author{Michela Geri}
\affiliation{Department of Materials Science and Engineering, Massachusetts Institute of Technology, Cambridge MA 02139, United States} 

\author{Brice Saint-Michel}
\affiliation{ENSL, CNRS, Laboratoire de Physique, F-69342 Lyon, France}
\affiliation{Present Address: Univ Gustave Eiffel, CNRS, Ecole des Ponts Paristech, UMR 8205 Laboratoire Navier, 5 Boulevard Descartes CEDEX 2, 77454 Marne-la-Vall\'ee, France}
    
\author{Thibaut Divoux}
    \affiliation{ENSL, CNRS, Laboratoire de Physique, F-69342 Lyon, France}

\author{Gareth H. McKinley}%
    \affiliation{Department of Mechanical Engineering, Massachusetts Institute of Technology, Cambridge MA 02139, United States}

\author{S\'ebastien Manneville}%
\email[]{sebastien.manneville@ens-lyon.fr      }
    \affiliation{ENSL, CNRS, Laboratoire de Physique, F-69342 Lyon, France}
\affiliation{Institut Universitaire de France}%

\date{\today}

\begin{abstract}
We study the local dynamics of a thixotropic yield stress fluid that shows a pronounced non-monotonic flow curve. This mechanically unstable behavior is generally not observable from standard rheometry tests, resulting in a stress plateau that stems from the coexistence of a flowing band with an unyielded region below a critical shear rate $\dot \gamma_c$. Combining ultrasound velocimetry with standard rheometry, we discover an original shear-banding scenario in the decreasing branch of the flow curve of model paraffin gels, in which the flow profile of the flowing band is set by the applied shear rate $\gd$ instead of $\dot \gamma_c$. As a consequence, the material slips at the walls with a velocity that shows a non-trivial dependence on the applied shear rate. To capture our observations, we propose a differential version of the so-called lever rule, describing the extent of the flowing band and the evolution of wall slip with shear rate. This phenomenological model holds down to very low shear rates, at which the dimension of the flowing band becomes comparable to the size of the wax particles that constitute the gel microstructure, leading to cooperative effects. Our approach provides a framework where constraints imposed in the classical shear-banding scenario can be relaxed, with wall slip acting as an additional degree of freedom.
\end{abstract}

\maketitle

\section{Introduction}

\begin{figure}
\centering
	\includegraphics[trim=0cm 0 0cm 0, clip, width=\linewidth]{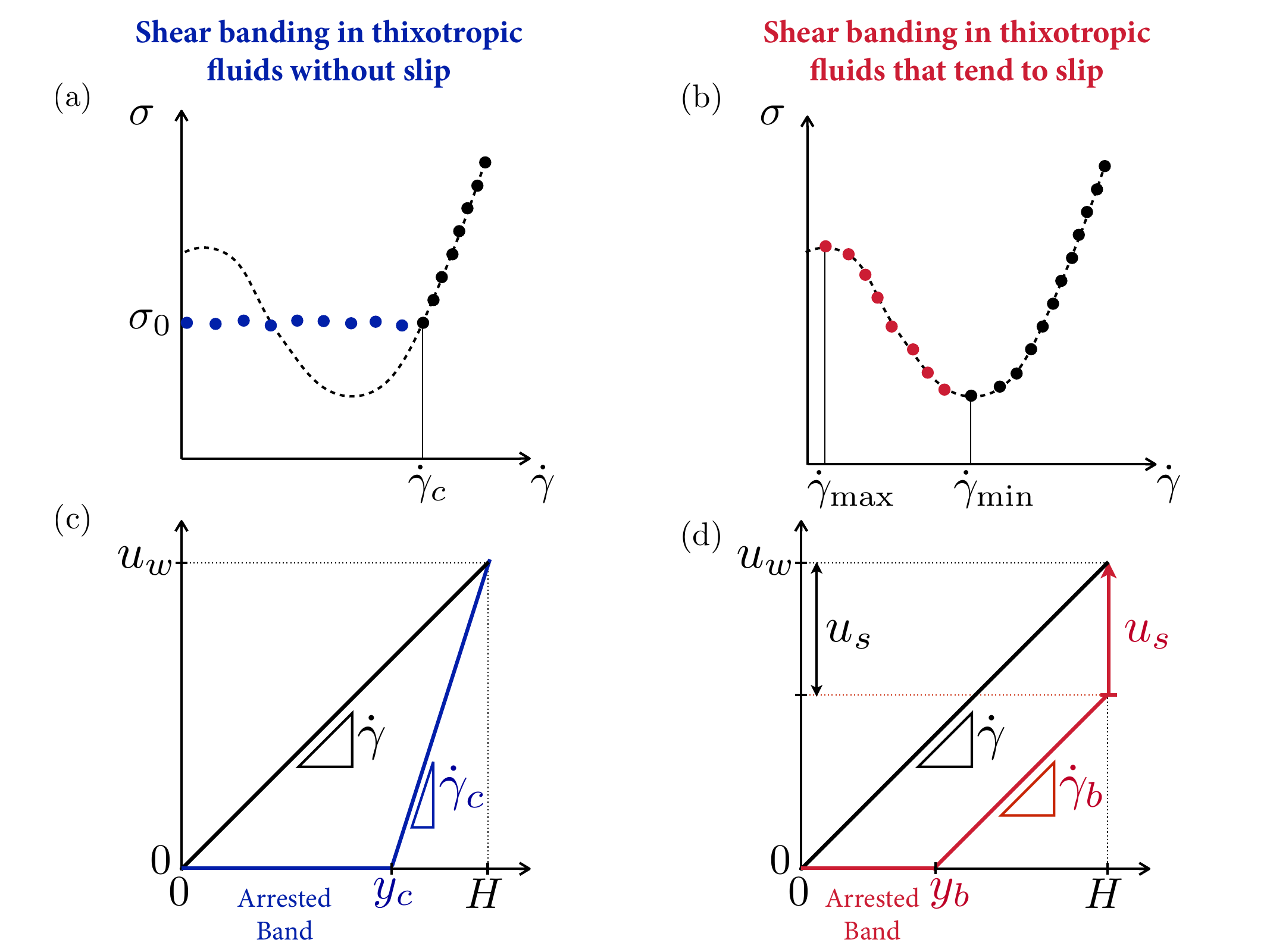}%
	\caption{(a) Schematic of the measured flow curve (solid circles) and underlying non-monotonic constitutive relation (dotted line) for thixotropic yield stress fluids that do not slip, and (c) corresponding velocity profiles for a generic point at $\dot \gamma < \gd_c$, where $u_w$ denotes the velocity of the moving wall, $H$ is the width of the gap, $y_c$ is the width of the arrested band and $\gd_c$ is the critical shear rate.
 (b)~Schematic for the scenario reported in model paraffin gels, which display an evident and measurable non-monotonic flow curve. The decreasing branch (colored in red) is observed for $\dot \gamma < \gd_\text{min}$, where $\gd_\text{min}$ corresponds to the local minimum in the flow curve. The velocity profile for a generic point with shear rate $\dot \gamma < \gd_\text{min}$ is shown in panel (d), where the arrested band width is $y_b$, the local shear rate in the flowing band is $\gd_b$ and $u_s$ is the slip velocity at the rotor.}
	{\label{fig:fig0}}
\end{figure}

Yield stress fluids (YSF) encompass a broad range of materials, from cosmetics and food products to cement pastes and waxy crude oils. These materials have in common a jammed or percolated microstructure conferring upon them solid-like properties under low external stresses. However, for stresses above a critical value, the microstructure yields, and the material flows. \cite{Bonn2017} This shear-induced solid-to-liquid transition often results in a time-dependent mechanical response referred to as thixotropy and leads to complex flow features, including stick-slip, fracture, and more generally, heterogeneous flow profiles. \cite{Persello1994,Divoux2010,Coussot2014,Divoux2016,Andrade2019} The latter phenomenon, coined \textit{shear banding}, has been mainly attributed to the competition between aging (often due to attractive interactions between constituents) and the rejuvenation imposed by external shear, \cite{Viasnoff2002,Cloitre2000,Bonn2002a} defining a critical shear rate $\gd_c$, below which a homogeneous shear flow becomes unstable.\cite{Coussot2002b,Fielding2007}
As sketched in Fig.~\ref{fig:fig0}(a), this competition is epitomized by a decreasing branch in the constitutive curve that relates stress $\sigma$ to shear rate $\dot\gamma$. It can be formally shown that fluid flow is unstable in this region \cite{Tanner1992,Yerushalmi1970} and therefore shear banding is always expected. Several experimental results show that for thixotropic YSF, when $\gd < \gd_c$
the local flow profile splits into a fluidized region sheared at a critical rate $\gd_c$ and an unyielded band, whose extent $y_c$ follows the \emph{lever rule}, i.e., grows proportionally to $(\gd_c - \dot \gamma)$ as the shear rate $\gd$ is decreased progressively below $\gd_c$. \cite{Coussot2002b,Manneville2008,Ovarlez2009,Fall2010,Fielding2014,Divoux2016} A schematic of this velocity profile is shown in Fig.~\ref{fig:fig0}(c) for a generic imposed shear rate $\gd < \gd_c$.
This phenomenology, which we refer to as \emph{classical shear-banding scenario}, is further characterized by the emergence of a stress plateau $\sigma_0$ in the flow curve (stress vs. shear rate) measured in rheological experiments. \cite{Divoux2016} In YSF, $\sigma_0$ has to be smaller than the static yield stress such that it can be sustained in both the unyielded and the flowing band.\cite{Martin2012} In the case of viscoelastic fluids, it has been shown that the selected stress is unique and independent of the band properties (see, e.g., Ref.~\onlinecite{Olmsted2008} and references therein). However, the picture is less clear for YSF, with experimental and numerical evidence of non-uniqueness of steady-state shear-banding features. \cite{Martin2012,Mohtaschemi2014,Divoux2016}

The phenomenology observed in the classical shear-banding scenario was further shown to be affected by boundary conditions and by the existence of wall slip. Originally described as a mere artifact that can be suppressed by well-chosen surface properties, wall slip appears to play a deeper, fundamental role in the flow properties of YSF.\cite{Buscall2010a,Aral1994,Cloitre2017,Derzsi2017,Malkin2018,He2019b} For instance, wall slip may affect the yielding transition and the steady-state flow properties of YSF.\cite{Gibaud2008,Gibaud2009,Cloitre2017} Nonetheless, the consensus remains that wall slip and the appropriate boundary conditions, more generally, can always be tuned independently of the bulk material rheological response.\cite{Mansard2014}

In this article, we investigate in detail the interplay between wall slip and shear banding in thixotropic YSF through experiments on model paraffin gels that are known to display pronounced non-monotonic flow curves under steady shear. \cite{Dimitriou2014,Mendes2015,Mendes2015a,Geri2017}
A schematic of the measured flow curve is shown in Fig.~\ref{fig:fig0}(b). Local velocity measurements coupled to standard rheometry allow us to unravel an original steady-state shear-banding scenario: along the decreasing branch of the flow curve, the shear rate in the fluidized band is controlled by the externally imposed shear rate $\dot \gamma$, rather than being fixed to $\dot \gamma_c$, as sketched in Fig.~\ref{fig:fig0}(d). Consequently, the slip velocity of the fluid at the wall displays a non-trivial dependence with $\gd$.
We propose a simple phenomenological model that accurately describes our experimental data, in which wall slip is directly related to the material bulk behavior, based on a differential formulation of the lever rule. This approach remains valid down to very low shear rates, at which cooperative effects arise as the dimension of the flowing band becomes comparable to the average particle size. \cite{Goyon2008,Bocquet2009,Goyon2010} Our experimental results on a specific thixotropic YSF also suggest a generalized approach to the classical lever rule that relaxes the constraints imposed in the classical shear-banding scenario and that highlights the importance of thixotropic time scales in determining the local dynamics of YSF.

\begin{figure*}[tb]
    \centering
	\includegraphics[width=0.75\textwidth]{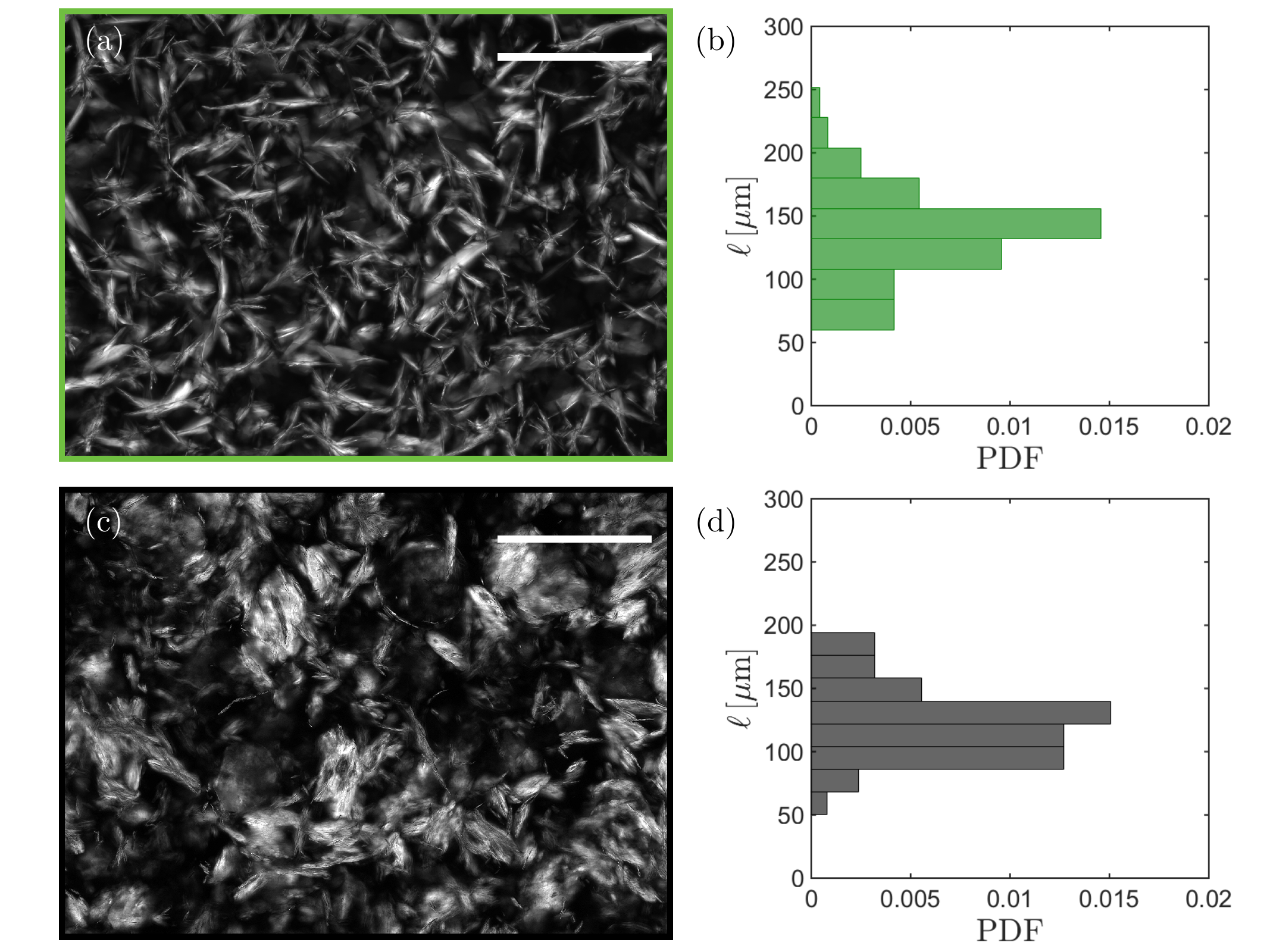}%
	\caption{(a) Birefringence image of a $\tenpc$ paraffin gel cooled under static conditions at the cooling rate of $\dot{T} = -0.2 \degCmin$ used in the rheo-velocimetry experiments. (b)~Probability density function (PDF) of the length $\ell$ of the paraffin crystallites obtained from image analysis of several birefringence images. (c)~Birefringence image of a $\tenpc$ paraffin gel cooled under dynamic conditions at the cooling rate ($\dot{T} = -0.2 \degCmin$) and sheared at $\gd = 50 \invs$ used in the rheo-velocimetry experiments. (d)~Corresponding probability density function (PDF) of the crystallite length $\ell$. Scale bars in (a) and (b) are both $300 \micron$. The average size of the crystallites is largely unaffected by shear and we measure $\langle \ell \rangle \simeq 130 \micron$ in both cases.}
	{\label{fig:figureSI_microscopy}}
\end{figure*}

\section{Materials and Methods}

\subsection{Sample preparation}

Model paraffin gels are prepared as described in Ref.~\onlinecite{Dimitriou2014} by dissolving a commercial wax made of linear $n$-paraffin chains (Sigma Aldrich $327212$, ASTM D $87$, melting point between $58^\circ \mathrm{C}$ and $62^\circ \mathrm{C}$) in heavy mineral oil (Sigma Aldrich, $330760$) of viscosity $\mu_o\simeq 100$~mPa.s at the working temperature of $26 \degC$. Solid wax is weighed at room temperature, then added to the liquid mineral oil, and the mixture is continuously stirred at high temperature ($T = 80 \degC$) overnight.
We prepare two large batches containing respectively $\tenpc$ and $\fifteenpc$ of paraffin wax, which are used for all subsequent experiments.

Before each experiment, the whole batch is heated back to $T = 80 \degC$ on a hot stirring plate to make sure that all paraffin chains are fully dissolved in the oil; then, the desired amount of sample is poured inside a Taylor-Couette geometry right before starting the cooling ramp (see Section~\ref{sec:rheo} below for the detailed protocol). Under these conditions, well above the wax appearance temperature ($T_{wa}$) for both the $\tenpc$ and $\fifteenpc$ concentrations, the samples are purely Newtonian at the beginning of each experiment, thus erasing any memory of their previous thermal or shear history. 

Upon cooling below $T_{wa}$, the paraffin chains precipitate out of solution and crystallize, forming discotic platelets with lateral dimensions that strongly depend on the cooling rate, final temperature and, less prominently, on the shear rate.\cite{Venkatesan2005} The typical thickness of each platelet is $\lesssim 1 \micron$ and is known to correspond to the average carbon number of the paraffin chains in solution.\cite{Smith1953,Singh2000} Note that, for these gels, temperature (and more specifically, the subcooling with respect to $T_{wa}$) is the main driving force in the crystallization process, not shear.\cite{Smith1953,Singh2000,Harris2023,Werner2023} From previous experiments, it is known that crystallization terminates relatively quickly once a temperature below $T_{wa}$ is reached.\cite{Geri2017}. Therefore, changes in the microstructure at the final working temperature only depend on the rearrangements of the already crystallized paraffin particles.

\subsection{Polarized light microscopy of paraffin gels}

Since paraffin crystallites are birefringent, we use polarized light microscopy to visualize the gel microstructure. A series of images is recorded at room temperature using an ABRIO (CRi, Inc.) camera and software on a Nikon TE-2000U inverted microscope using a $20\times$ objective. Two types of preparation protocols are employed in order to compare the morphology of single wax crystals in the case of static cooling or dynamic cooling (i.e., cooling while also imposing shear).

\begin{figure*}[tb]
	\includegraphics[trim=0cm 0 0cm 0, clip, width=0.9\textwidth]{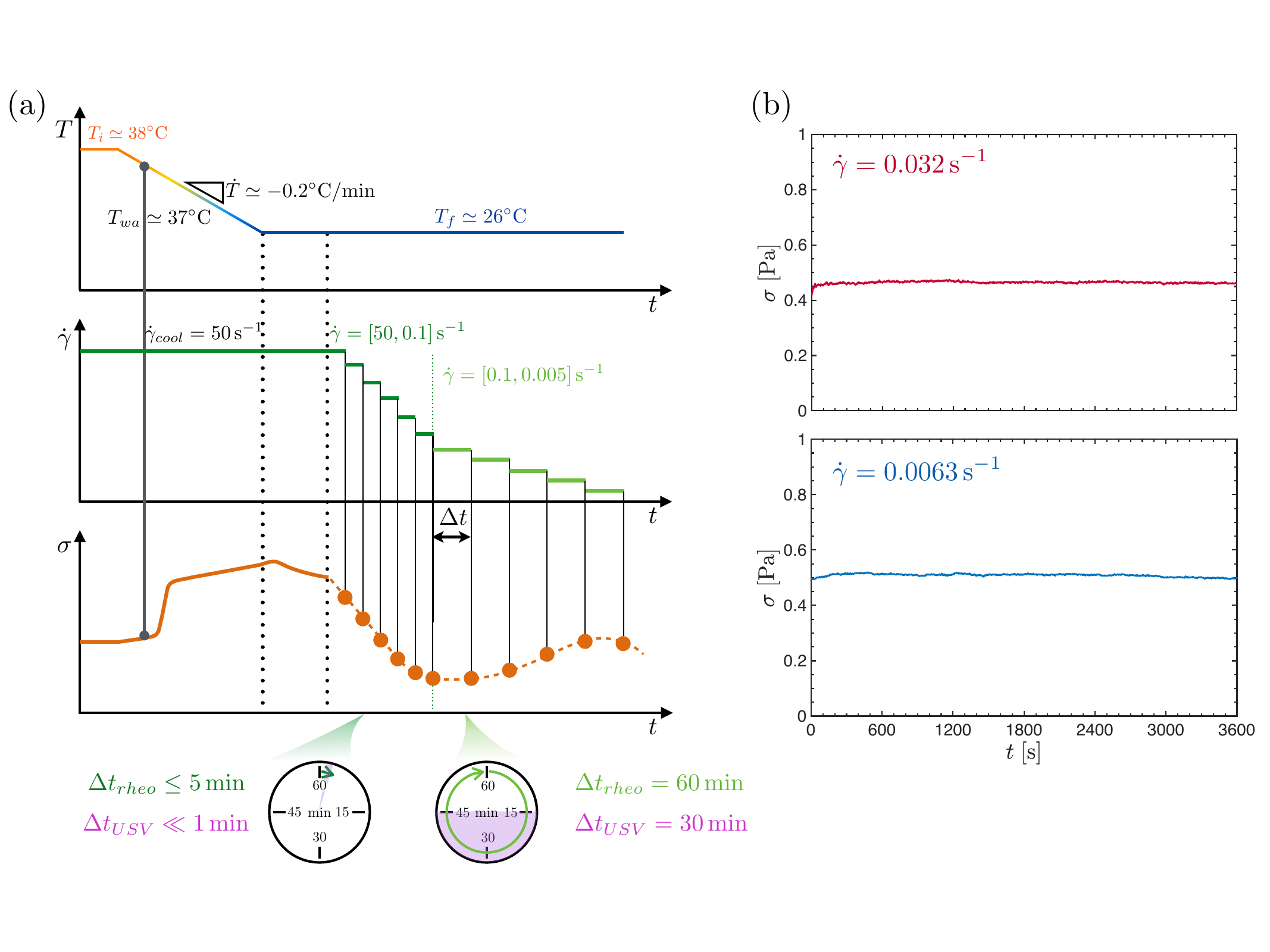}%
	\caption{(a)~Experimental protocol used for rheo-velocimetry experiments. From top to bottom: temperature $T$ imposed using an external chiller connected to the water tank surrounding the Taylor-Couette cell, shear rate $\gd$ imposed by the rheometer and corresponding stress response $\sigma$ together with a schematic of the acquisition times $\Delta t_{\rm rheo}$ and $\Delta t_{\rm USV}$  used respectively for rheological measurements and for ultrasonic velocimetry in different ranges of shear rates.
	(b)~Stress response $\sigma(t)$ recorded in the \tenpc\ gel for shear rates $\gd=0.032 \invs$ (top) and $0.0063 \invs$ (bottom), applied during the flow curve of Fig.~\ref{fig:figure1}(a).}
	{\label{fig:figureSI_Protocol}}
\end{figure*}

For the static case, a small visualization chamber is made by laying three stripes (approximately $3 \times 20\, \mathrm{mm}$) of Parafilm on a glass slide ($75 \times 25\, \mathrm{mm}$). A cover glass ($22 \times 22\, \mathrm{mm}$) is then sealed on top of the Parafilm by heating the slide on a hot plate at about $80^\circ \mathrm{C}$, which makes the Parafilm adhesive. Next, the liquid paraffin--oil mixture is pipetted in between the bottom slide and the cover glass at the same temperature of $80^\circ \mathrm{C}$. Finally, the cell is quickly placed on a Peltier stage to impose the same cooling rate as that used in the rheo-velocimetry experiments, namely $\dot{T} = -0.2 \degCmin$, until room temperature is reached.

For the dynamic case, the melted gel sample is spread on a glass slide ($75 \times 25\, \mathrm{mm}$) in contact with the Peltier stage of a stress-controlled rheometer (DHR-3, TA Instruments) heated to $80^\circ \mathrm{C}$. Shear is applied using a parallel-plate geometry of diameter $20 \mm$ covered with sandpaper (similar roughness as in the velocimetry experiments) and with a gap of $1 \mm$. The sample is then cooled under a constant shear rate $\gd = 50 \invs$ at a cooling rate $\dot{T} = -0.2 \degCmin$ until room temperature is reached. At the end of the process, we raise the head of the rheometer and transfer the glass slide directly onto the microscope for visualization.

Images representative of both cooling processes are reported in Fig.~\ref{fig:figureSI_microscopy}(a) for the static case and Fig.~\ref{fig:figureSI_microscopy}(c) for the dynamic case. Although the platelet-like crystals are arranged very differently in the two cases, most probably due to the flow generated when raising the shearing tool in the last step of dynamic cooling, a quantitative analysis of a series of images shows very similar distributions of the crystallite characteristic in-plane size $\ell$ [see Fig.~\ref{fig:figureSI_microscopy}(b,d)], with an average value of approximately $130 \micron$ in both cases. This suggests that the crystallite in-plane size is not influenced by the continuous shear during crystallization at this very low cooling rate. As mentioned above, the thickness of each platelet is well below $1 \micron$, as expected from the average carbon number of the paraffin mixture used.\cite{Singh2000}

To summarize, the microstructure of paraffin gels consists of platelet-like paraffin particles of characteristic size $\ell\simeq 130 \micron$, primarily set by the cooling rate, and thickness below $1 \micron$. Crystalline bridges as well as steric interaction among these discotic particles give rise to an elastic network and to a complex nonlinear response under shear.\cite{Singh2000,Miyazaki2014} 

\subsection{Rheological protocol}
\label{sec:rheo}

The rheological behavior of the same paraffin gels employed in this study has already been thoroughly investigated in Ref.~\onlinecite{Geri2017}. Other works on similar systems also exist in the literature \cite{Dimitriou2014,Mendes2015,Mendes2015a} and generally show a similar picture that identifies these gels as thixotropic yield stress fluids with a pronounced non-monotonic flow curve and tendency to slip. Here, we focus specifically on the local dynamics underlying this non-monotonic flow curve, and aim at understanding how flow heterogeneities may develop such that the decreasing branch is experimentally accessible.
In order to proceed with the mechanical characterization of our model paraffin gels under shear, each solution is loaded into a homemade Taylor-Couette cell (or concentric-cylinder shear cell) mounted on a stress-controlled rheometer (AR-G2, TA Instruments) equipped with ultrasonic velocimetry as described in Ref.~\onlinecite{Gallot2013}. We use the feedback loop of the rheometer to impose a global shear rate $\gd$ and measure the corresponding shear stress $\sigma$ as a function of time. The shear cell
consists of a fixed cylindrical cup made of sandblasted PMMA and of an inner rotating cylinder (bob) made of polyoxymethylene (Delrin\textsuperscript{\textregistered{}}) covered with sandpaper of surface roughness $35 \micron$. The radius of the bob is $R_i=23.9 \mm$, and its height is 65~mm. The (inner) radius of the cup is $R_o=25 \mm$ so that the working gap is $H = R_o-R_i =1.1 \mm$. The curvature of the cylindrical geometry leads to a relative decrease of the shear stress from the bob to the cup given by $\delta \sigma/\sigma=(R_o/R_i)^2-1$. This can be easily derived by solving the Cauchy momentum equation for Stokes flow ($\bm{\nabla} \cdot \bm{\sigma} = \textbf{0}$) in cylindrical coordinates.\cite{Divoux2016} With the present geometry, $\delta \sigma/ \sigma \simeq 0.09$, where $\sigma = \sigma_{r\theta}$ is the only non-zero shear stress component. We can therefore conclude that the effect of curvature is negligible and assume a quasi-homogeneous stress field across the gap, in contrast to other recent work on similar waxy suspensions,\cite{Andrade2020} where $R_o/R_i\simeq 2.7$ or $\delta \sigma / \sigma \simeq 6.3$. Moreover, we deliberately choose a large roughness, which is usually supposed to minimize wall slip, to highlight the fact that wall slip may occur in spite of rough boundary conditions. 

Figure~\ref{fig:figureSI_Protocol}(a) depicts the combined cooling and rheological protocol imposed in our experiments. Due to the thermal sensitivity of the probe used for ultrasonic velocimetry, the water bath surrounding the Taylor-Couette cell and the probe must be kept at temperatures lower than $40 \degC$. Therefore, the water bath is first stabilized at $T_i = 38 \degC$, while the bob is pre-heated at about $50 \degC$, before pouring the paraffin--oil mixture heated at $T = 80 \degC$ in the Taylor-Couette cell. The sample is subsequently sheared at $\dot{\gamma}_{\rm cool} = 50 \invs$, starting right after pouring and during the whole cooling process. Once the sample temperature equilibrates at $T_i$, as measured by a hand-held K-type thermocouple, the chiller of the water bath is set to cool the sample at a rate $\dot{T} = -0.20 \pm 0.05 ^\circ \mathrm{C/min}$ until a final temperature $T_f = 26 \degC$. After reaching $T_f$, the mixture is sheared for an additional $20 \min$. Finally, the flow curve is measured by progressively decreasing the shear rate down to $0.005 \invs$. The shear rate is swept down logarithmically with 10 points per decade between $50\invs$ and $1\invs$ and 20 points per decade between $1\invs$ and $0.1\invs$. For each applied shear rate above $0.1\invs$, the rheometer waits for the stress to vary by less than 5\% over $90 \s$ (with a time limit per point set to $5 \min$) before decreasing the shear rate to the next value. From $\gd = 0.1\invs$ down to $0.005\invs$, a constant shear rate is imposed over $\Delta t_{\rm rheo}=60 \min$. As seen in Fig.~\ref{fig:figureSI_Protocol}(b) for the $\tenpc$ paraffin gel, this protocol allows us to ensure that all initial transients have died out even at the lowest shear rates, while accommodating for the time $\Delta t_{\rm USV}$ needed for acquiring ultrasound velocimetry maps at steady state, as explained in the next section. A similar level of stationarity and residual fluctuations is observed in all stress responses for the $\fifteenpc$ paraffin gel.

\subsection{Local velocity measurements}

As explained in full detail in Ref.~\onlinecite{Gallot2013}, our ultrafast ultrasound imaging technique outputs maps of the tangential velocity component $u_\theta(y,z,t)$ as a function of the radial distance $y$ to the outer cup, the vertical position $z$ along the ultrasonic probe, and time $t$. Such velocity maps result from the cross-correlation of successive ultrasound images of the material under shear, recorded from the emission and back-scattering of plane pulses sent with a repetition frequency that is inversely proportional to the applied shear rate. In the case of the present paraffin gels with wax concentrations $\tenpc$ and $\fifteenpc$, the wax crystallites scatter ultrasound efficiently enough that ultrasound images can be obtained directly, without requiring any seeding of the samples by acoustic contrast agents. The duration $\Delta t_{\rm USV}$ of ultrasound image acquisition ranges from a few seconds at the highest shear rate to about 30 minutes at $\gd = 0.005\invs$.

\begin{figure}[tb]
	\includegraphics[width=\linewidth]{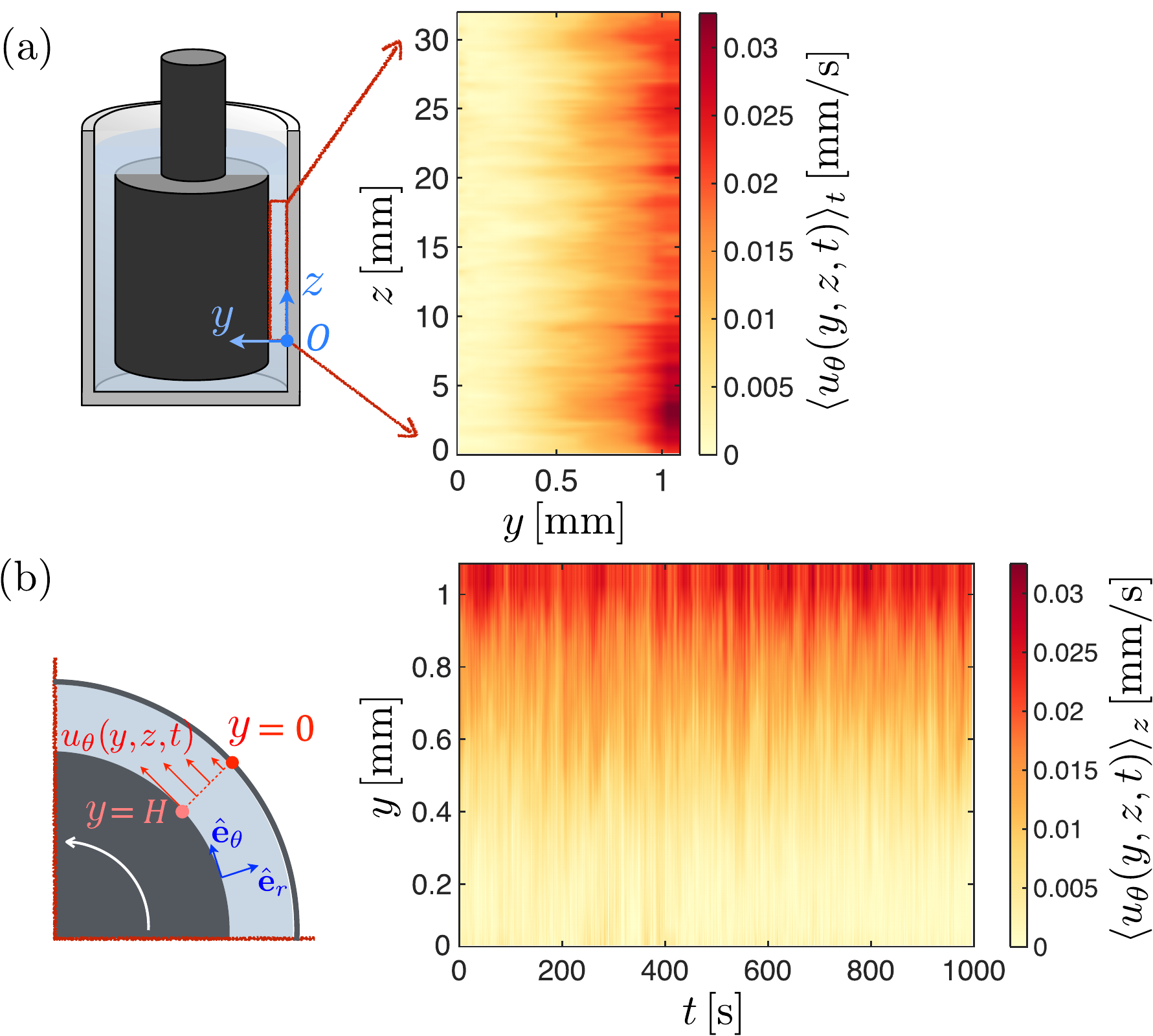}%
	\caption{Time-resolved velocity field $u_\theta(y,z,t)$ recorded in the $\tenpc$ paraffin gel under an imposed shear rate $\gd = 0.025 \invs$. (a)~Time-averaged velocity map $\langle u_\theta(y,z,t) \rangle_t$ taken over the whole duration of the ultrasound acquisition. (b)~Space-time diagram of the velocity data $\langle u_\theta(y,z,t) \rangle_z$ averaged over the vertical direction $z$.}
	{\label{fig:figureSI_spaceTime}}
\end{figure}

Figure~\ref{fig:figureSI_spaceTime} shows an example of velocity data recorded on the $\tenpc$ paraffin gel for $\gd=0.025 \invs$. The origins $y=0$ and $z=0$ are taken respectively at the fixed wall and at the bottom of the ultrasonic probe. As seen in Fig.~\ref{fig:figureSI_spaceTime}(a), the velocity field $\langle u_\theta(y,z,t) \rangle_t$ averaged over $t$ for the whole acquisition window of duration $\Delta t_{\rm USV}=10^3$~s does not show any significant variation along the vertical direction $z$. Similarly, the velocity $\langle u_\theta(y,z,t)\rangle_z$ averaged over $z$ and presented as a function of $y$ and $t$ in the spatiotemporal diagram of Fig.~\ref{fig:figureSI_spaceTime}(b) fluctuates around a mean without any systematic trend over $10^3$~s. Analogous maps are produced for all other shear rates under study as well as for the $\fifteenpc$ paraffin gel, which indicates that the flow is stationary and invariant along the vertical direction $z$. This allows us to focus on the velocity profiles $u(y)\equiv\langle u_\theta(y,z,t) \rangle_{z,t}$ averaged both in time and along the vertical direction, and look for flow heterogeneity along the velocity gradient direction $y$ only.

In the following, we shall see that the averaged velocity profiles $u(y)$ may combine wall slip and shear banding, in which a non-flowing, unyielded region coexists with a flowing band as sketched in Fig.~\ref{fig:fig0}(d). To quantify the average local shear rate $\gd_b$ in the flowing band, the extent of the unyielded band $y_b$ and the slip velocity $u_s$, we consider the time-resolved data $u(y,t)\equiv\langle u_\theta(y,z,t) \rangle_{z}$. For each time $t$, $u(y,t)$ is fitted to a linear profile in the two bands (with zero slope for the unyielded band). The time-averaged slope of each fitted flowing band provides an estimate for $\gd_b$ and the time-averaged intercept between the two linear profiles an estimate for $y_b$, while the slip velocity at the moving wall is extracted by taking the time-averaged difference between the wall velocity and the velocity calculated based on the fitted linear profile extrapolated at the wall. The velocity at the fixed wall is always found to be negligible up to experimental uncertainty, although velocities closest to the cup may be overestimated due to the signal analysis used to remove fixed ultrasonic echoes.\cite{Gallot2013} Error bars in all velocity profiles and calculated quantities are computed based on standard deviations over time and are, therefore, characteristic of the temporal fluctuations observed experimentally.

\section{Results} 

\subsection{Evidence for wall slip and shear banding}

\begin{figure}[tb]
	\includegraphics[trim=6cm 1.5cm 6cm 1cm, clip,width=\columnwidth]{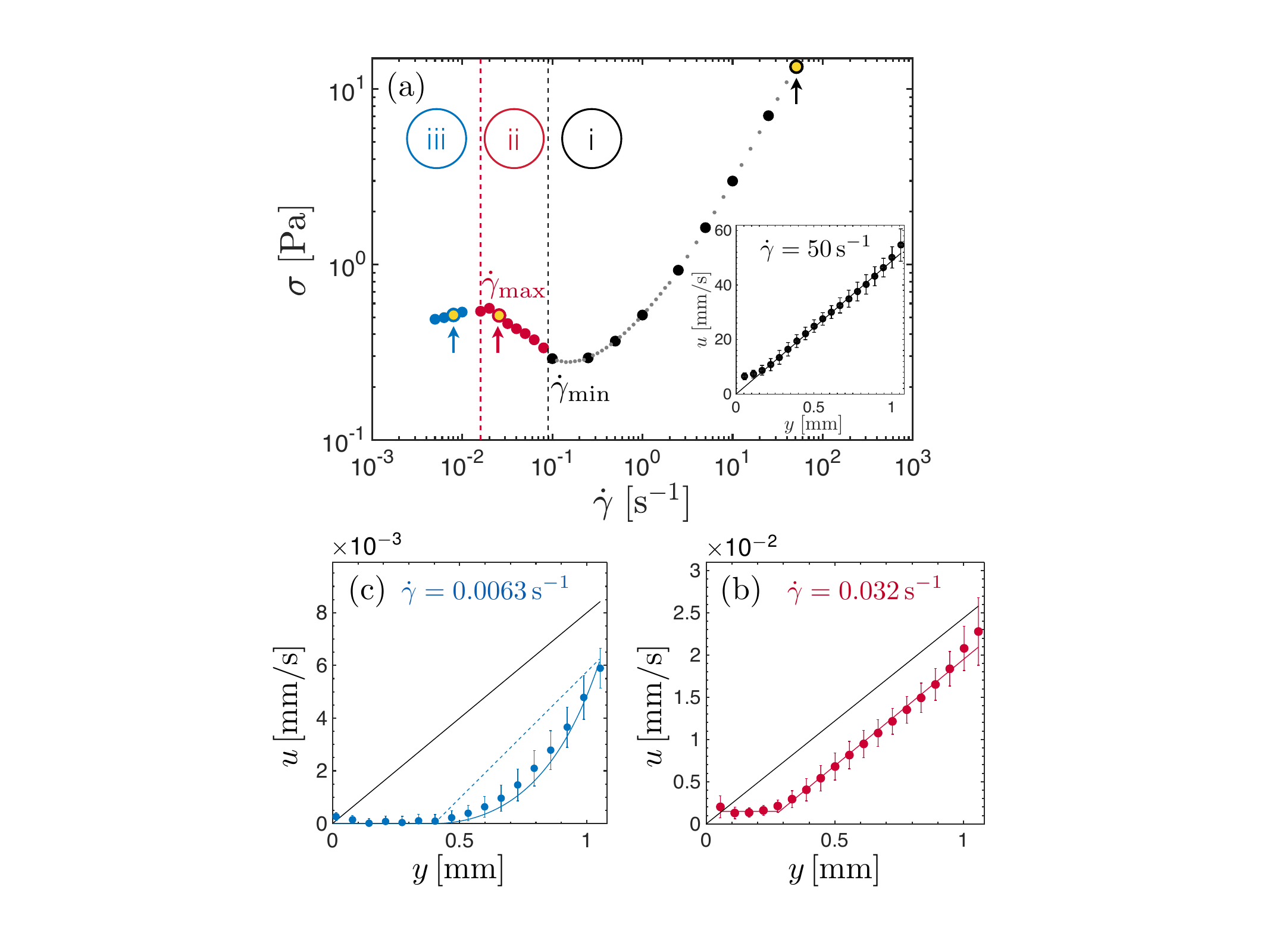}%
	\caption{Rheological response of the $\tenpc$ paraffin gel: (a)~Flow curve of shear stress $\sigma$ vs. shear rate $\dot \gamma$. Velocity profiles $u(y)$ recorded simultaneously delineate three regimes: ($i$)~homogeneous velocity profile (see inset for profile measured at $\gd = 50 \invs$); ($ii)$~banded velocity profile with distinct wall slip at the rotating bob as illustrated in (b) for $\gd = 0.032 \invs$; ($iii$)~curved velocity profile as illustrated in (c) for $\gd = 0.0063 \invs$. The coordinate $y$ denotes the distance from the fixed wall of the Taylor-Couette cell with gap $H=1.1 \mm$. Black lines are the homogeneous velocity profiles expected without wall slip. Colored solid lines in (b,c) are best fits to the velocity profiles, while the dashed line in (c) shows the linear profile that would be expected from the average shear rate in the sheared band.}
	{\label{fig:figure1}}
\end{figure}

Figure~\ref{fig:figure1}(a) illustrates the flow behavior of the $\tenpc$ paraffin gel, measured by a decreasing ramp of shear rate steps. The flow curve shows a pronounced decreasing branch delimited by two specific shear rates: $\gd_{\rm min} \simeq 0.09 \invs$ at which the stress reaches a local minimum, and $\gd_{\rm max} \simeq 0.016 \invs$ at which the stress reaches a weak local maximum. 
These two specific shear rates separate three different flow behaviors as confirmed by velocity profiles recorded simultaneously to the flow curve. For large shear rates, i.e., $\gd\ge \gd_{\rm min}$ (regime \circled{\regI}), the flow curve monotonically increases with shear rate, the corresponding velocity profiles are homogeneous, and wall slip remains negligible as expected for rough walls [see inset of Fig.~\ref{fig:figure1}(a)]. Intermediate shear rates, i.e., $\gd_{\rm max} \le \gd \le \gd_{\rm min}$ (regime \circled{\regII}), correspond to the decreasing branch. The base flow is unstable and the velocity profiles separate into two regions, a flowing shear band close to the inner rotating bob (located at $y=1.1 \mm$) and an unyielded, solid-like region close to the fixed outer cup [see Fig.~\ref{fig:figure1}(b) for $\gd = 0.032 \invs$]. The spatial extent $y_b$ of the unyielded band increases as the shear rate decreases, while the average local shear rate in the flowing band $\gd_b$ remains very close to the global imposed shear rate $\gd$. Concomitantly, we measure noticeable slippage of the flowing band at the bob with a slip velocity $u_{s}$. Finally, for $\gd \le \gd_{\rm max}$ (regime \circled{\regIII}), the quiescent band takes up about half the gap and the velocity profile in the flowing band shows increasing curvature, while $\gd_b$ (which here represents the average shear rate in the flowing band) almost matches $\gd$ [see Fig.~\ref{fig:figure1}(c) for $\gd = 0.0063 \invs$]. Note that for shear rates in regime \circled{\regIII}, instead of using a linear fit, the curved velocity profile in the flowing band is fitted for $y\in[y_b,H]$ by a polynomial of order 3, $u(y)=\sum_{n=0}^3a_n y^n$ with $a_i$ the polynomial coefficients. From these, $\gd_b$ is obtained as the \emph{average} local shear rate, i.e., $\gd_b=\frac{1}{h}\int_{y_b}^H \frac{\di u}{\di y}(y)\di y=\frac{1}{h}\sum_{n=1}^3 n a_n \int_{y_b}^H y^{n-1}\di y$, with $h=H-y_b$. The striking curvature of $u(y)$ in regime \circled{\regIII} is specifically discussed below in Sect.~\ref{sec:coop}. 

Moreover, we emphasize that, based on the radial distribution of the ultrasonic signal within the gap, which is directly correlated to the local concentration of scatterers (i.e., paraffin crystallites), we do not detect any clear difference in particle distribution between the flowing and quiescent band either in regime \circled{\regII} or in regime \circled{\regIII}. Therefore, a scenario based on flow--concentration coupling, as suggested to explain non-classical shear banding in some hard-sphere colloidal glasses,\cite{Ballesta2008,Besseling2010} is unlikely to hold in our rather dilute gels.

\begin{figure}[tb]
\centering
	\includegraphics[trim=8.6cm 0cm 8.3cm 1cm, clip, width=0.75\columnwidth]{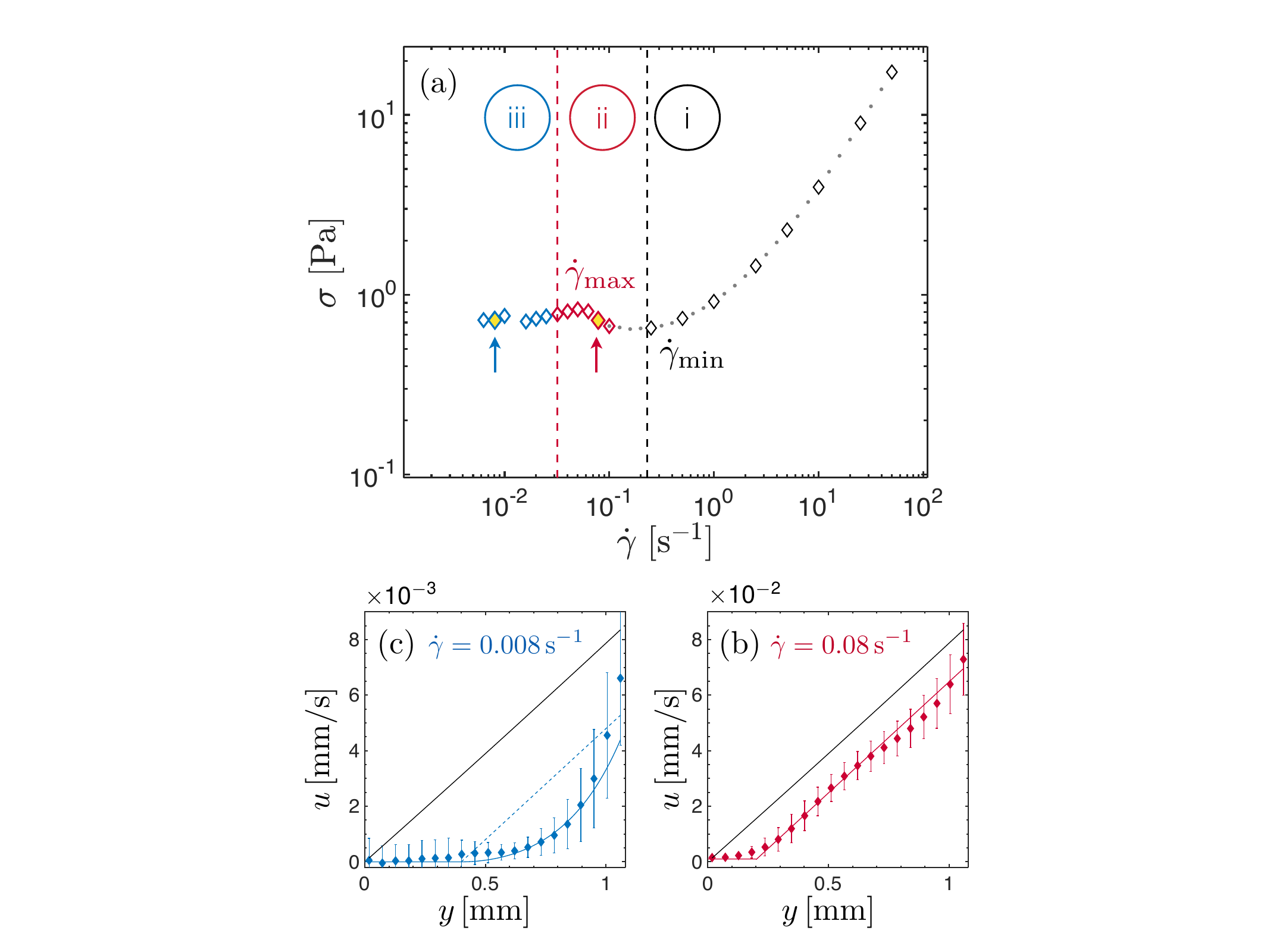}
	\caption{Rheological response of the $\fifteenpc$ paraffin gel: (a) Flow curve $\sigma$ vs. $\gd$. Banded velocity profiles for (b) $\gd=0.08 \invs$ and (c) $\gd=0.008 \invs$. Black lines in (b) and (c) are the homogeneous velocity profiles expected without wall slip. Colored solid lines are best fits to the velocity profiles, while the dashed line in (c) shows the linear profile that would be expected from the average shear rate in the sheared band.}
	{\label{fig:figureSI_15pc}}
\end{figure}

In order to test for the robustness of the previous observations with respect to the wax content, we investigate the $\fifteenpc$ gel with the same protocol. From Fig.~\ref{fig:figureSI_15pc}, it is clear that the $\fifteenpc$ gel follows the very same phenomenology, although with larger values of $\gd_{\rm min}$ and $\gd_{\rm max}$ than for the $\tenpc$ gel. We observe the same regimes in the flow curve [Fig.~\ref{fig:figureSI_15pc}(a)], which correspond to the same types of velocity profiles [Fig.~\ref{fig:figureSI_15pc}(b,c)] as in Fig.~\ref{fig:figure1}.

\begin{figure}[tb]
	\includegraphics[trim= 5.9cm 0cm 4.5cm 0,clip,width=\columnwidth]{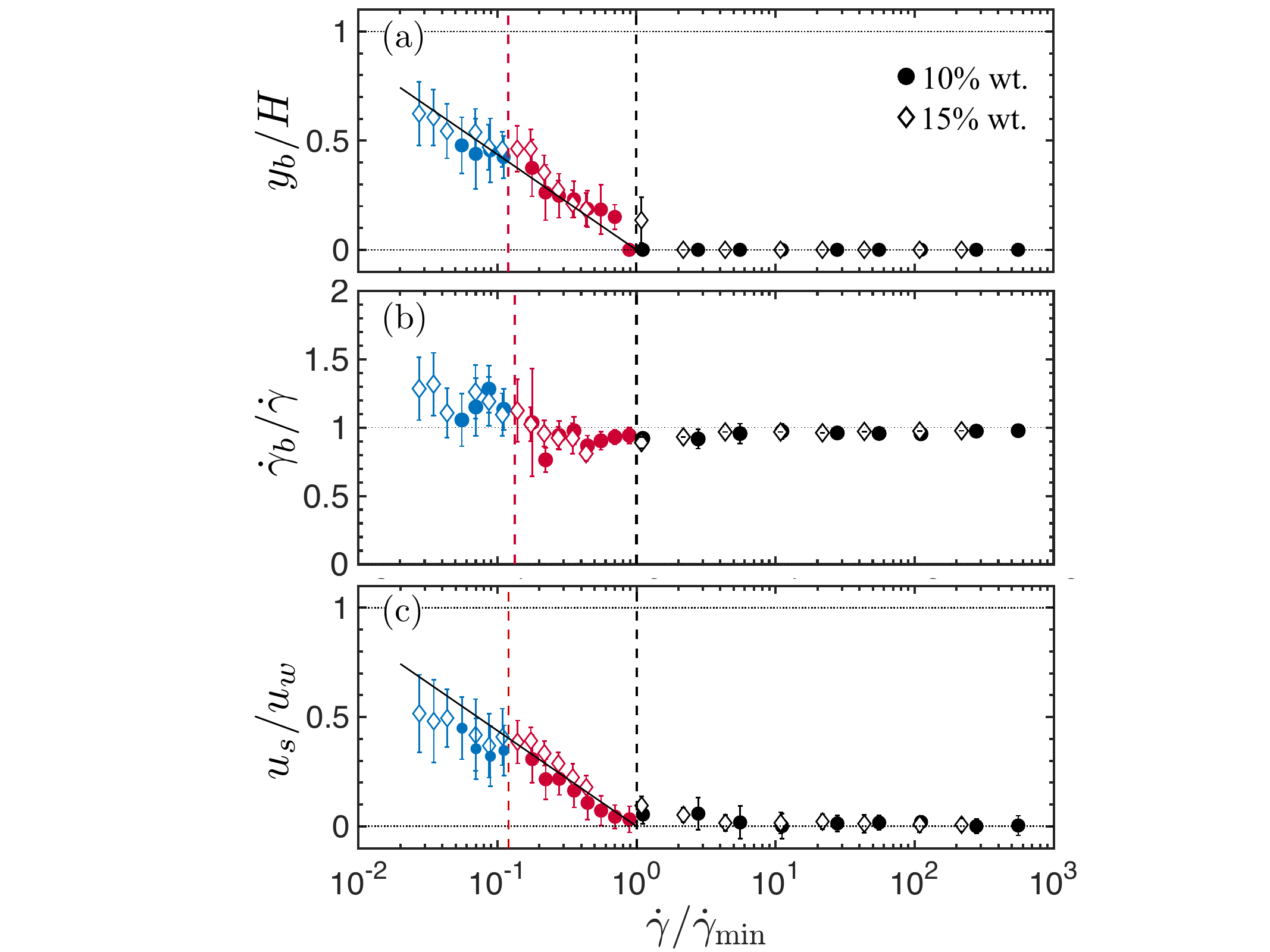}%
	\caption{(a) Lateral extent $y_b$ of the solid-like band across the gap, (b)~local shear rate in the flowing band $\gd_b$ and (c)~slip velocity $u_s$ (made dimensionless respectively by the gap width $H$, the imposed shear rate $\gd$ and the wall velocity $u_w=\gd H$). Data for both the $\tenpc$ (filled circles) and $\fifteenpc$ (open diamonds) gels are presented as logarithmic functions of the dimensionless shear rate $\gd/\gd_{\rm min}$, with $\gd_{\rm min} = 0.09 \invs$ and $0.23 \invs$ respectively, as extracted from the flow curves. Solid lines are fits to Eq.~\eqref{eq:USV_yb_beta} with $\beta=0.19$.}
	{\label{fig:figure2}}
\end{figure}

In Fig.~\ref{fig:figure2}, we further analyze the velocity profiles as a function of $\gd/\gd_{\rm min}$ for both wax concentrations. Remarkably, all three observables defined so far and recalled in Fig.~\ref{fig:figure2}(d), namely, the normalized extent of the solid-like band, $y_b/H$ [Fig.~\ref{fig:figure2}(a)], the normalized average local shear rate in the flowing band, $\gd_b/\gd$ [Fig.~\ref{fig:figure2}(b)] and the ratio $u_{s}/u_{w}$ of the slip velocity $u_{s}$ at the rotating bob to the bob velocity $u_{w}$ [Fig.~\ref{fig:figure2}(c)], collapse on the same master curves. This suggests a generic underlying physical mechanism, in which both the unyielded region and the relative slip velocity grow for decreasing shear rates below $\gd_{\rm min}$, while the flowing band always experiences the global imposed shear rate $\gd$. Finally, we emphasize that the fact that no wall slip is detected in regime \circled{\regI}, even very close to $\gd_{\rm min}$, confirms the effectiveness of the sandpaper in suppressing slip whenever a homogeneous velocity profile is observed [see Fig.~\ref{fig:figure1}(a)]. We checked that the shear-banding phenomenology reported here is robust and also observed when both surfaces are sandblasted or smooth, although experiments performed without sandpaper clearly show a significant additional slip velocity even in regime \circled{\regI}. In the next section, we model the interplay between wall slip and shear banding in terms of a generalized lever rule that allows us to account for the observed bulk behavior.

\subsection{Phenomenological modeling based on time-scale separation}

\begin{figure}
	\includegraphics[trim= 0.75cm 2.5cm 0.75cm 2.5cm,clip,width=0.9\columnwidth]{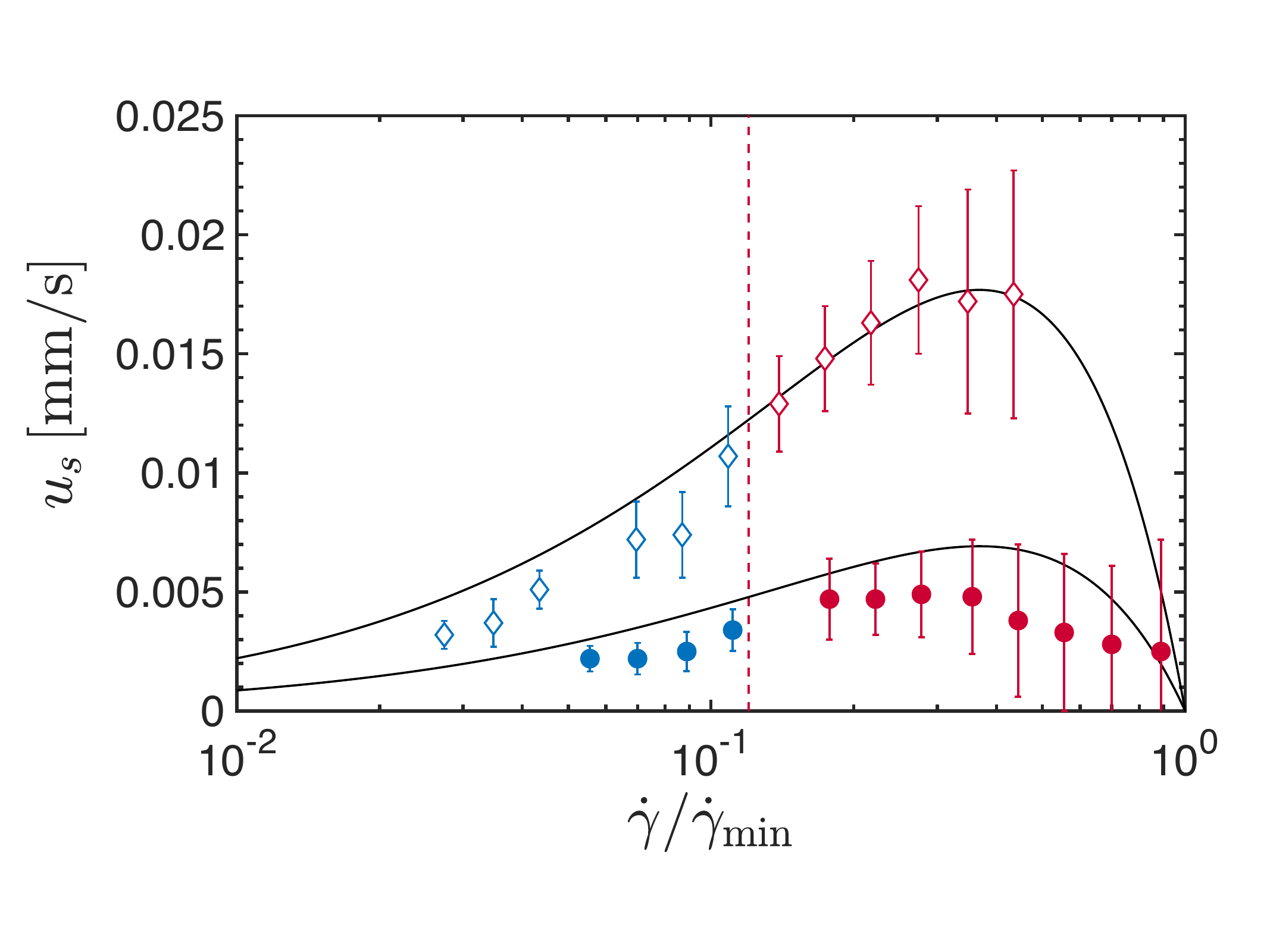}%
	\caption{Slip velocity $u_s$ as a function of the dimensionless shear rate $\gd/\gd_{\rm min}$ in regimes \protect\circled{\regII} and \protect\circled{\regIII} for the $\tenpc$ (filled circles, $\gd_{\rm min} = 0.09 \invs$) and $\fifteenpc$ (open diamonds, $\gd_{\rm min} =0.23 \invs$) gels. The predictions of Eq.~\eqref{eq:USV_yb_beta} for both gel concentrations with $\beta = 0.19$ are shown by the black solid lines without any additional fitting parameter.}
	{\label{fig:figure3}}
\end{figure}

The above experimental 
observations starkly contrast with previous reports on steady-state shear banding, in which the flowing band experiences shear at the critical shear rate $\gd_c$ and the ``classical'' lever rule for the extent of the shear band, $y_b\propto (\gd_c - \gd)$, holds for $\gd<\gd_c$. As recalled in the introduction, these features have been associated with a non-monotonic underlying constitutive behavior, appearing as a stress plateau at $\sigma_0$ in experimental flow curves below $\gd_c$.\cite{Coussot2002b,Fielding2007} By contrast, in our experiments, the extent of the solid-like band $y_b$ displays a nonlinear, logarithmic dependence on $\gd$ [Fig.~\ref{fig:figure2}(a)] and the average local shear rate in the flowing band $\gd_b$ coincides (within error bars) with the externally applied shear rate $\gd$ [Fig.~\ref{fig:figure2}(b)]. Moreover, the fact that strong wall slip accompanies shear-banded flows suggests that slippage is key to accounting for the steady-state shear banding we observe. Indeed, as shown in Fig.~\ref{fig:figure3}, the computed slip velocities $u_s$ display non-trivial trends as a function of $\gd$ for both paraffin concentrations.

In order to rationalize our observations, we now revisit the premises of the classical lever rule. First, we note that the classical lever rule assumes the continuity of the velocity profile \emph{in the absence of wall slip}. \cite{Manneville2008,Ovarlez2009,Divoux2016} Second, thixotropic YSF obeying the lever rule are usually modeled using a structural parameter possessing a single characteristic time scale $\tau$ driving the dynamics of both the yield stress at rest and the plastic viscosity under flow.\cite{Coussot2002a,Ragouilliaux2006,Ovarlez2009,Cheddadi2012} Such a time scale governs the evolution equation of an internal parameter that describes the level of \textit{build up} or \textit{break down} of the fluid microstructure.\cite{Barnes1997,Mewis2009,Larson2015,Larson2019} However, extensive experiments have shown that model paraffin gels can build up a yield stress very fast, while their \emph{plastic} viscosity changes over a much longer time scale.\cite{Dimitriou2013,Mendes2015,Mendes2015a,Geri2017,Andrade2020} Modeling this class of thixotropic YSF requires the introduction of \textit{two} microstructural parameters and therefore \textit{two} characteristic time scales,\cite{Geri2017} denoted respectively $\tau_y$ (over which the yield stress builds up in the unyielded region) and $\tau_v$ (over which the plastic viscosity changes in the flowing band), with $\tau_{y} \ll \tau_{v}$. For a similar chemical composition, we have already shown in Ref.~\onlinecite{Geri2017} that fits to an elastoviscoplastic model of thixotropic YSF lead to $\tau_v\simeq 460$~s and $\tau_y\simeq$~10~s. Note that the longer time scale $\tau_v$ is much smaller than the one-hour duration of each step at a given shear rate, which allows us to consider that a steady state is reached for all $\gd$ in regime \circled{\regII}.

Based on these observations, we hypothesize that when the time-scale separation $\tau_{y} \ll \tau_{v}$ holds, the flowing band cannot adjust its plastic viscosity fast enough to a fixed critical shear rate, here denoted $\gd_{\rm min}$. As a consequence, the local shear rate $\gd_b$ remains close to the global imposed shear rate $\gd$, while the fluid preferentially starts to {\it slip} at the boundaries of the shear cell through a thin lubricating layer mainly composed of solvent.

Within this framework, we can further derive expressions for the extent of the solid-like region $y_b$ and the slip velocity $u_s$ as a function of the external imposed shear rate.
Note that mass conservation ($\bm{\nabla} \cdot \bm{u} = \textbf{0}$) requires $u_\theta = u(y)$, in agreement with experimental observations [see Fig.~4(a)], while the simplified Cauchy momentum equation yields an approximately constant uniform stress $\sigma$ across the narrow gap of the Couette cell, as discussed in Sect.~\ref{sec:rheo}. We start by ensuring the continuity of the velocity profile through $\gd_b h + u_s = u_w$, where $u_w$ is the wall velocity [see Fig.~\ref{fig:fig0}(d)]. We further consider the slip layer as an additional band of finite thickness $\delta_{s}$, with the same viscosity $\mu_o$ as the suspending oil. Given the stress homogeneity, we can estimate the slip velocity as $u_s=\sigma \delta_{s}/\mu_o$.
Perturbing the relation $\gd_b h + u_s = \gd H$ leads to: 
\begin{equation}
\label{eq:differential}
h \,\di \gd_b + \gd_b\, \di h + \di\!\left( \frac{\sigma \delta_{s}}{\mu_o} \right) = H \,\di \gd.
\end{equation}
To exactly solve Eq.~\eqref{eq:differential}, it is necessary to couple it with an explicit, temporally- and spatially-resolved constitutive relation for the paraffin gel. However, much progress can be obtained analytically by using scaling arguments alone. Independently of the details of the constitutive model, since $\tau_{y} \ll \tau_{v}$, any microstructural change within the flowing band develops much more slowly than in the unyielded band, where the yield stress is building up. Hence, for any given time interval $\di t$, we may assume $h \di \gd_b \ll \gd_b \di h$ and, up to first order, we may neglect the contribution from $h \di \gd_b$.
Of the remaining terms, we can see that all of them are, dimensionally, a shear rate multiplied by a length. Concerning shear rates, $\gd_b \simeq \gd$, while $\sigma/\mu_o \sim \gd \mu_p/\mu_o < 10 \gd$, where $\mu_p/\mu_o < 10$ is the ratio between the plastic viscosity of the gel and the oil viscosity. However, concerning lengths, $\delta_{s}=\mu_o u_s/\sigma\simeq 1$--5~$\mu$m is smaller than both $h$ and $H$ by two to three orders of magnitude, as already reported in other yield stress materials.\cite{Salmon2003b,Salmon2003c,Meeker2004,Zhang2017,Pemeja2019} Therefore, we conclude that the term related to wall slip may also be neglected in Eq.~\eqref{eq:differential}. Finally, based on our experimental observations, we may further substitute $\gd_b = \gd$, which leads to the simple following expression of the lever rule in differential form:
\begin{equation} \label{eq:RevisedLeverRule}
\gd \di h  \simeq H \di \gd \,.   
\end{equation} 
Upon integration of Eq.~(\ref{eq:RevisedLeverRule}) from $\gd_{\rm min}$ at which $h=H$ (i.e., the entire material is homogeneously sheared) to $\gd<\gd_{\rm min}$ at which $h=H-y_b$, we obtain:
	\begin{equation}
	\label{eq:USV_yb_beta}
		\frac{y_b}{H} = - \beta \ln \left( \frac{\gd}{\gd_{\rm min}} \right) = \frac{u_s}{u_w} \,,
	\end{equation}	
where $\beta \sim \mathcal{O}(1)$ is a dimensionless scaling factor accounting for every simplification leading to Eq.~(\ref{eq:RevisedLeverRule}). The last equality results from velocity continuity, imposing $u_s=\gd_b y_b=\gd y_b$. Figure~\ref{fig:figure2}(a) shows that Eq.~\eqref{eq:USV_yb_beta} accurately describes the normalized extent of the unyielded region $y_b/H$ and the normalized slip velocity $u_s/u_w$ with $\beta = 0.19\pm 0.05$ in \emph{both} regimes \circled{\regII} and \circled{\regIII} [see black solid lines in Fig.~\ref{fig:figure2}(a,c)]. Interestingly, the same value of $\beta$ fits both data sets equally well, confirming the consistency and robustness of the modified lever rule through \textit{a priori} independent wall slip and bulk flow measurements. Finally, with $u_w=\gd H$, Eq.~\eqref{eq:USV_yb_beta} reads $u_s=\beta H \gd\ln(\gd_{\rm min}/\gd)$, reproducing very well our experimental wall slip data [Fig.~\ref{fig:figure3}].
In particular, the non-monotonic evolution of $u_s$ with $\gd$ appears as a distinctive consequence of the modified lever rule.

\subsection{Flow cooperativity at very low shear rates}
\label{sec:coop}

\begin{figure}[tb]
	\includegraphics[trim=1.3cm 5.3cm 0cm 5.5cm, clip, width=0.9\columnwidth]{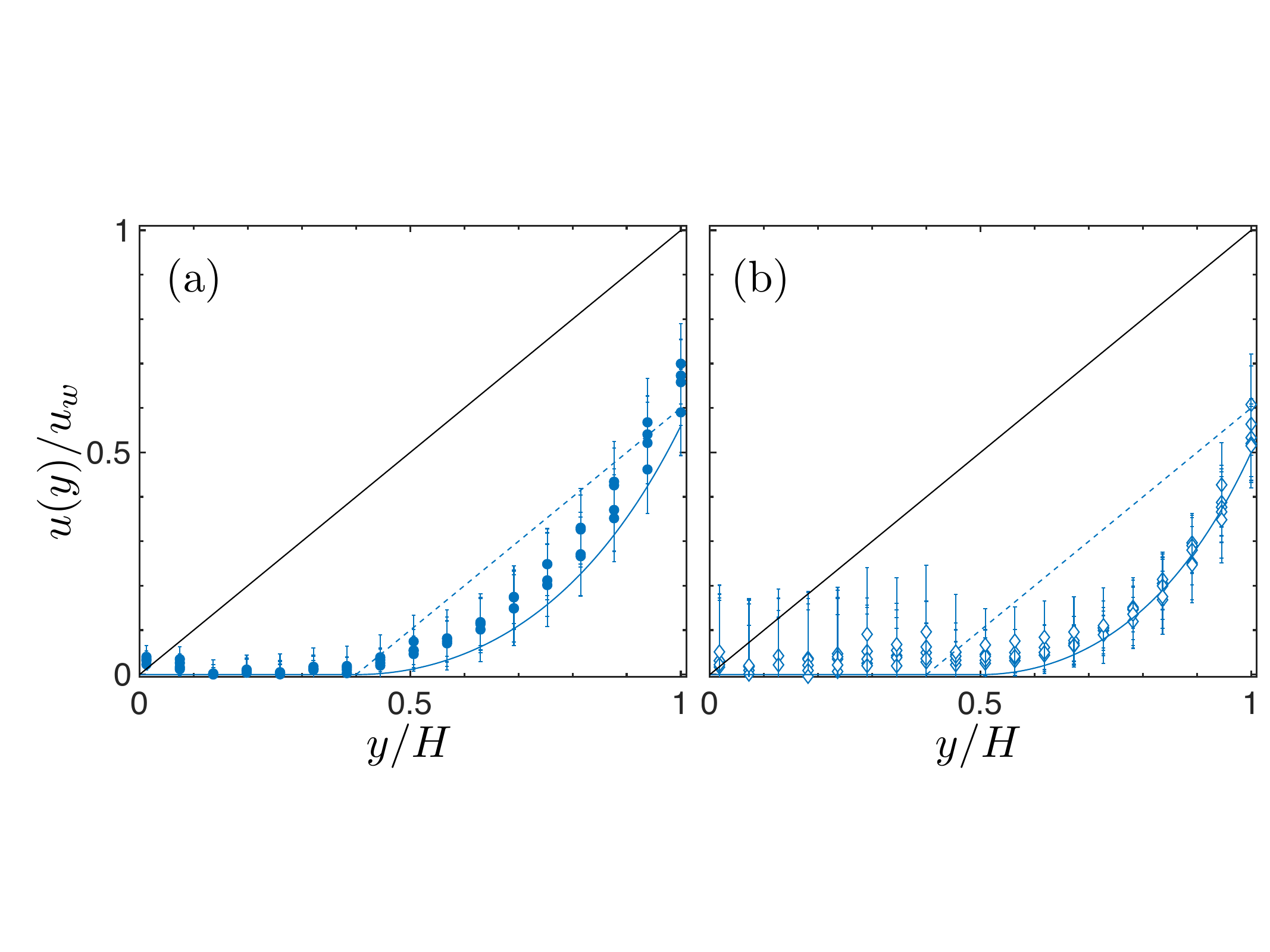}%
	\caption{Normalized velocity profiles measured over regime \protect\circled{\regIII} (a)~in the $\tenpc$ paraffin gel and (b)~in the $\fifteenpc$ paraffin gel. Standard deviations are reflective of the temporal fluctuations and of experimental uncertainty on our ultrasound velocity measurements at very low shear rates. Solid black lines show the theoretical profile for a Newtonian fluid in the absence of wall slip; blue solid lines are fits to the nonlocal profile in Eq.~\eqref{eq:USV_vel_prof_regIII_dim}; blue dashed lines are the profiles for a shear rate fixed at the measured average value.}
	{\label{fig:figure4}}
\end{figure}

Figure~\ref{fig:figure4} gathers the normalized velocity profiles $u(y)/u_w$ recorded in regime \circled{\regIII} for both the $\tenpc$ and $\fifteenpc$ paraffin gels. When compared to the velocity profiles in regime \circled{\regII} [see Figs.~\ref{fig:figure1}(b) and \ref{fig:figureSI_15pc}(b)], it is clear that the velocity profiles in the flowing region show a much more pronounced curvature in regime \circled{\regIII}, recalling the profiles observed in confined suspensions when spatial cooperativity is present\cite{Seth2012,Goyon2008, Bocquet2009} or in granular flows.\cite{Pouliquen2009,Kamrin2012} Inspired by these previous results, we solve a nonlocal equation for the local plastic fluidity $f(y) = \gd(y)/\sigma$ to derive an expression for the velocity profile in the flowing band, i.e., for $y_b \le y \le H$. This nonlocal equation reads
\begin{equation}
\label{eq:fluidity}
\xi^2 \frac{\de^2 f }{\de y^2} = f -f_b\,,
\end{equation}
where $\xi$ is the flow cooperativity length and $f_b$ is the bulk fluidity. In our case, since the material outside the flowing band is an unyielded viscoplastic solid, we impose $f_b = f(y_b) \equiv 0$ and define $f (H) \equiv f_w > 0$ at the moving wall, which leads to:
\begin{equation}
f(y) = f_w \frac{\sinh[(y-y_b)/\xi]}{\sinh(h/\xi)}\,,
\end{equation}
with $h=H-y_b$. Further integrating the local kinematic relationship $\frac{\di u(y)}{\di y} \equiv \gd(y)= f(y) \sigma$ and imposing $u(y_b) = 0$ yields: 
	\begin{align}
		\label{eq:USV_vel_prof_regIII_dim}
		u(y) =u(H) \frac{\cosh[(y-y_b)/\xi] - 1}{\cosh(h/\xi) - 1}\,,
	\end{align}
where $u(H) = \xi f_w \sigma\frac{\cosh(h/\xi) - 1}{\sinh(h/\xi)}$.
In Fig.~\ref{fig:figure4}, we fit the experimental data to Eq.~\eqref{eq:USV_vel_prof_regIII_dim} in order to determine an estimate of the cooperativity length $\xi$. In view of the large experimental uncertainty resulting from the very low velocity levels, we focus on the velocity profiles averaged over all shear rates available in regime \circled{\regIII}. Since $f_w$ is not known {\it a priori}, we also have to treat $u(H)$ as a fitting parameter. We find $\xi \simeq 200 \micron$, which is comparable to the average dimension of single wax crystallites, while $u(H)/u_{w} = 0.55$ and $y_b/H = 0.4$ for both gel concentrations, consistent with the measured values shown in Fig.~\ref{fig:figure2}. The agreement between theory (blue solid lines) and experiments suggests that nonlocal effects can indeed explain the pronounced curvature observed in the velocity profiles measured in regime \circled{\regIII}. Moreover, despite the confinement of the sample microstructure, the generalized lever rule still holds in regime \circled{\regIII}, provided one defines $\gd_b$ as the average of the local shear rate in the flowing band.

\section{Discussion and conclusion}

In this work, we have identified an original shear-banding scenario in model paraffin gels, in which the flow remains homogeneous with negligible wall slip along the increasing branch of the flow curve for $\gd>\gd_{\rm min}$ (regime \circled{\regI}), whereas it separates into two bands for $\gd<\gd_{\rm min}$ (regimes \circled{\regII} and \circled{\regIII}), one region being unyielded and solid-like close to the fixed boundary and the other flowing with a locally uniform shear rate $\gd_b=\gd$ and showing a large slip velocity at the moving wall. We have introduced a phenomenological version of the lever rule that allows us to model accurately the non-trivial relationship between the extent of the solid-like band $y_b$, the local shear rate in the flowing band $\gd_b$, and the slip velocity $u_s$, together with their dependence on the global imposed shear rate $\gd$.

The predictions of Eq.~\eqref{eq:USV_yb_beta} hold all along the decreasing branch of the flow curve, i.e., in regime \circled{\regII}, as well as in the limit of very low shear rates, i.e., in regime \circled{\regIII}. In the latter regime, the flowing band no longer results in a homogeneous shear rate $\gd_b=\gd$, but rather displays a strongly curved velocity profile. We interpret this curvature as the result of cooperative effects when the size of the flowing band becomes comparable to that of single wax particles.\cite{Goyon2008,Bocquet2009,Seth2012,Serial2021} In particular, a nonlocal model based on a diffusion equation for the fluidity  [Eq.~\eqref{eq:fluidity}] correctly fits the velocity profiles in the flowing band with a cooperativity length $\xi\simeq 200~\mu$m, consistent with the characteristic lateral dimension of paraffin platelets $\ell\simeq 130 \micron$.
The positive correlation between the characteristic length $\xi$ and the average particle size $\ell$, together with the observation that the measured velocity profiles are clearly curved only in region \circled{\regIII}, suggest that flowing aggregates within the shear band are made of stacked particles largely aligned in the flow direction. Geometrically, this is the only possible configuration that allows our platelet-like particle clusters to be small enough with respect to the gap. Stacked clusters have been reported before in simulations of non-Brownian suspensions \cite{Meng2008} and can be observed in the birefringence images of the paraffin gels investigated [see Fig.~\ref{fig:figureSI_microscopy}(c)], while alignment of anisotropic particles for sufficiently strong flows has been linked to hydrodynamic slip at the particle-fluid interface.\cite{Youngren1975,Kamal2020} When the shear rate is below $\dot{\gamma}_\text{max}$, shear alignment of the clusters is lost and the prevalence of rotation and tumbling \cite{Jeffery1922} translates into an effective hydrodynamic radius of the stacked agglomerates that approaches the in-plane platelet dimension $\ell$. As a consequence, nonlocal effects start to appear and dominate the flow behavior giving rise to the curved profiles observed in region \circled{\regIII}.

Note, however, that modeling our observations based solely on a nonlocal approach with a shear-rate dependent cooperativity length, which may diverge in the vicinity of yielding as in Ref.~\onlinecite{Kamrin2012}, would not account for the full richness of velocity profiles reported here. In particular, a modified lever rule of the form given by
Eq.~\eqref{eq:RevisedLeverRule} is necessary to account for our observation that $\gd_b=\gd$ in regime \circled{\regII}, and to quantitatively predict the evolution of the slip velocities reported in Fig.~\ref{fig:figure3}. Moreover, the condition $u_s \le u_w$ implies that Eq.~\eqref{eq:USV_yb_beta} should hold down to $\gd= \exp(-1/\beta)\gd_{\rm min} \simeq 0.0052\gd_{\rm min}$, which is approximately five times smaller than the minimum shear rate achieved in our velocimetry experiments. Therefore, future work using even lower imposed shear rates and longer acquisition times should assess whether Eq.~\eqref{eq:USV_yb_beta} remains valid deeper into regime \circled{\regIII}.

The phenomenological modeling approach proposed in this article quantitatively describes both wall slip and bulk flow heterogeneity in a thixotropic YSF for which the usual assumptions underpinning the classical lever rule do not hold. Still, our simple approach does not  constitute an alternative to the standard steady-state shear-banding scenario in the absence of wall slip. It rather {\it generalizes} the standard scenario: indeed, if $\tau_y \gtrsim \tau_v$ and no wall slip occurs, the local shear rate becomes $\gd_c$ and Eq.~(\ref{eq:RevisedLeverRule}) reduces to $\gd_c \di h = H \di \gd$, so that $\gd_c h_c = \gd H $ upon integration, which is the classical lever rule. Furthermore, a key observation is that a single value of the dimensionless coefficient $\beta$ in Eq.~\eqref{eq:USV_yb_beta} quantitatively describes the measured slip velocities in both the $\tenpc$ and $\fifteenpc$ paraffin gels. We may thus hypothesize that the value of $\beta$ is mostly controlled by the nature and the roughness of the shearing surfaces, while the value of $\gd_{\rm min}$ mostly depends on the system concentration. Future experiments, which would systematically vary the surface properties for a given system, are clearly needed to test the validity of such a hypothesis.

The time-scale separation $\tau_y \ll \tau_v$ and its microscopic origin appear as important yet often overlooked possible feature in many thixotropic YSF, for which it is often simply assumed that the two time scales coincide. In the case of the present paraffin gels, the highly anisotropic, platelet-like shape of the interacting microcrystals that make up the gel microstructure (see Fig.~\ref{fig:figureSI_microscopy}) is likely to account for the short time scale $\tau_y$ required to rebuild a locally-percolated microstructure with solid-like properties compared to the time scale $\tau_v$ for the evolution of the plastic viscosity.
In this framework, due to the dense packing of anisotropic microcrystals, we expect a severe increase in the time scale for the microstructure to orient under an external shear, and thus an increase of $\tau_v$ by analogy with suspensions of rigid fibers.\cite{Butler2018, Bounoua2019}
Hence, the decreasing branch of the flow curve, which is mechanically unstable and usually observed only in transient responses,\cite{Mas1994,Pignon1996,Grondin2008,Moller2008} becomes accessible here. From the present findings, we may anticipate that a large time-scale separation between $\tau_y$ and $\tau_v$ should generically lead to non-monotonic steady-state flow curves in other thixotropic YSF, and to the same combination of wall slip and a shear-banded velocity profile.
The next step is thus to investigate a broader range of experimental systems to validate the generality of the phenomenology reported here.

Moreover, although we have taken care of ensuring that {\it steady states} are reached in our experiments, it is important to recall that even small jumps in the shear rate may involve long transients in the stress response of thixotropic YSF. For instance, increasing the shear rate by steps is most often associated with stress overshoots.\cite{Dullaert2006,Mewis2009,Wei2019,Larson2019,Benzi2021a,Benzi2021b,Benzi2023} 
During such overshoots, the local maximum in shear stress may be larger than the maximum stress previously experienced by the sample and/or than the yield stress. If the flowing shear band were to adjust its local shear rate to a hypothetical $\gd_\text{c}$, the stress within the band would take a value that remains above yield for a time long enough to lead to the fluidization of the whole gap. 
The present scenario involving wall slip  appears as an alternative path for YSF with large time-scale separation to adjust to stress variations. However, our results raise the question of whether one scenario may be metastable relative to the other, and of the influence of the experimental protocol. In particular, one may ask whether a change in the time $\Delta t_{\rm rheo}$ spent at each shear rate, especially in the vicinity of $\gd_\text{min}$, would yield substantially different results. From what we observed in both rheometric and velocimetry data, when imposing a shear rate slightly higher or exactly equal to $\gd_\text{min}$, the sample reaches equilibrium very quickly, in about $2 \minutes$. Therefore, waiting longer than $\Delta t_{\rm rheo}=5 \minutes$ in regime \circled{\regI} should not significantly affect our observations. Still, additional experiments that systematically explore the influence of the flow protocol will have to be performed to confirm such a hypothesis.

The above discussion also prompts us to consider \textit{unsteady} heterogeneous flows under external shear. For instance, it is known that shear startup flows of thixotropic YSF and the corresponding yielding transition are accompanied by wall slip, even with rough boundary conditions.\cite{Gibaud2008,Divoux2011b,Grenard2014} During these short periods of time, YSF (even those with isotropic constituents) experience conditions similar to those reported in the present work, i.e., $\tau_y \ll \tau_v$, which strongly suggests that wall slip also acts as an external degree of freedom during transient flows. In this framework, Eq.~(\ref{eq:USV_yb_beta}) provides a functional form for the scaling of the slip velocity with the externally imposed shear rate $\gd$. This expression could be coupled to spatially-resolved models, e.g., soft glassy rheology or simpler fluidity models,\cite{Fielding2014} to account for wall slip in transient flows.  

Finally, our results at low shear rates, where cooperative effects dominate the flow profile, call for incorporating wall slip into spatially-resolved models to further investigate the interplay between wall slip and nonlocal effects. To date, spatially-resolved models with cooperative effects that successfully account for both steady-state and complex long-lasting transient flows \cite{Benzi2016,Nicolas2018,Benzi2019,Benzi2021a,Benzi2021b} do not include wall slip.
This theoretical effort is all the more needed, given that distinguishing experimentally between the impact of wall slip and  of cooperative effects on the flow profile is a major challenge, as both effects occur over comparable length scales.
The present experimental results should thus serve as a benchmark for testing such generalized theories in an effort to fully understand viscoplastic flow processes under very small imposed shear rates close to the yielding transition. More broadly, the observations and accompanying model of shear banding and wall slip provide a broader framework for understanding the complexity of heterogeneous flows of thixotropic YSF that are of interest in engineering, geophysical and biomedical applications.

\section*{Author Contributions}
CRediT: MG and BSM: conceptualization, data curation, formal analysis, investigation, methodology, software, visualization, writing-original draft; TD, GHM, and SM: conceptualization, formal analysis, funding acquisition, methodology, project administration, resources, supervision, validation, writing- review \& editing.

\section*{Conflicts of interest}
There are no conflicts to declare.

\section*{Acknowledgements}
We are very thankful to Dr.~Bavand Keshavarz for machining the Couette rotor with great precision. We thank the MIT-France program for supporting this collaboration between MIT and ENS Lyon. MG and GHM are also grateful to Chevron ETC and the MIT Energy Initiative for supporting portions of this research.





%

\end{document}